\documentclass{article}
\usepackage{waspaa21,amsmath,graphicx,url,times}
\usepackage{color}
\usepackage{algorithmic}
\usepackage{graphicx}
\usepackage{subcaption}
\usepackage{textcomp}
\usepackage{xcolor}
\usepackage{lipsum}
\usepackage{cases}
\usepackage{multirow}
\usepackage{comment}
\usepackage[T1]{fontenc}

\title{SALADnet: Self-Attentive multisource Localization \\ in the Ambisonics Domain}
\name{Pierre-Amaury Grumiaux,$^{1}$
      Sr\dj{}an Kiti\'c,$^{1}$
      Prerak Srivastava,$^{2}$
      Laurent Girin,$^{3}$
      Alexandre Gu{\'e}rin$^{1}$}
\address{$^1$Orange Labs, Cesson-S{\'e}vign{\'e}, France\\ 
         $^2$Univ. de Lorraine, Inria, Nancy, France\\
         $^3$Univ. Grenoble Alpes, GIPSA-lab, Grenoble-INP, CNRS, Grenoble, France}

\begin{document}

\ninept
\maketitle

\begin{sloppy}

\vspace{-2cm}
\begin{abstract}

In this work, we propose a novel self-attention based neural network for robust multi-speaker localization from Ambisonics recordings. Starting from a state-of-the-art convolutional recurrent neural network, we investigate the benefit of replacing the recurrent layers by self-attention encoders, inherited from the Transformer architecture. We evaluate these models on synthetic and real-world data, with up to 3 simultaneous speakers. The obtained results indicate that the majority of the proposed architectures either perform on par, or outperform the CRNN baseline, especially in the multisource scenario. Moreover, by avoiding the recurrent layers, the proposed models lend themselves to parallel computing, which is shown to produce considerable savings in execution time.

\end{abstract}

\begin{keywords} Sound source localization, neural networks, self-attention, Ambisonics, parallel computing.

\end{keywords}

\vspace{-0.2cm}
\section{Introduction}
\label{sec:intro}

Sound source localization (SSL) is an active research field with various applications such as source separation \cite{chazan_multi-microphone_2019}, speech recognition \cite{lee_dnn-based_2016} or human-robot interaction \cite{li_reverberant_2016}. Traditional methods to address SSL are based on eigenvalue decomposition of a multichannel signal covariance matrix \cite{schmidt_multiple_1986}, time-difference of arrival (TDOA) estimation \cite{knapp_generalized_1976}, or beamforming \cite{dmochowski_generalized_2007}. These methods, and their variants, enjoy widespread popularity, but are known to lack robustness in noisy and reverberant environments.

More recently, machine learning methods have greatly improved the performance of SSL systems. In particular, deep neural networks have been proposed to improve single-source \cite{vesperini_neural_2016}, as well as multi-source DOA estimation \cite{adavanne_direction_2018,perotin_crnn-based_2019}. They were shown to improve the robustness of SSL in challenging conditions compared to traditional methods. Among deep learning models, different architectures have been proposed: convolutional neural networks (CNNs) \cite{vera-diaz_towards_2018}, convolutional recurrent neural networks (CRNNs) \cite{ adavanne_direction_2018,perotin_crnn-based_2019}, U-net architectures \cite{chazan_multi-microphone_2019}, autoencoders (AEs) \cite{liu_direction--arrival_2018} or attention-based neural networks \cite{schymura_exploiting_2021}. Also, various types of input features have been used, such as raw signal waveforms \cite{vera-diaz_towards_2018}, features based on the short-time Fourier transform (STFT) \cite{yalta_sound_2017}, correlation-based features \cite{vesperini_neural_2016, liu_direction--arrival_2018}, Ambisonics features \cite{perotin_crnn-based_2019, adavanne_direction_2018}, or combinations of different input features \cite{xue_sound_2020, cao_two-stage_2019}. These neural network-based systems function either in regression \cite{vesperini_neural_2016, vera-diaz_towards_2018, schymura_exploiting_2021} or in classification mode \cite{perotin_crnn-based_2019, adavanne_direction_2018, liu_direction--arrival_2018}. On the one hand, regression-based SSL provides DOA estimates with any arbitrary resolution, at the expense of dealing with the well-known source permutation problem in the case of multiple sources localization \cite{subramanian_deep_2021}. On the other hand, classification-based SSL directly provides (multiple) DOA estimates on a discrete spatial grid. 

When considering multiple sound source localization for Ambisonic signals, the CRNN-based neural networks perform particularly well, as reported in \emph{e.g.} \cite{adavanne_direction_2018,perotin_crnn-based_2019,krause_comparison_2021, grumiaux_improved_2021}. The major downside of such models is their sequential nature, \emph{i.e.} their recurrent layers cannot be efficiently parallelized. 
However, these recurrent layers are important for modeling the temporal dynamics, therefore, it is suboptimal to have them simply discarded or replaced by convolutional layers, as it it leads to significant drop in localization accuracy \cite{krause_comparison_2021,perotin_crnn-based_2019}. Beyond computational concerns, another potential issue is that the recurrent layers are essentially based on the first order Markov model, meaning that their output depends only on the hidden state and the current input, hence they do not directly incorporate information from the other elements of a sequence. 

Recently, the \emph{attention mechanism} has emerged as a promising alternative to model temporal dependencies. Originally proposed in \cite{bahdanau_neural_2016} in conjunction with recurrent neural networks (RNNs) for natural language processing (NLP), attention efficiently learns the interdependencies of elements (\emph{e.g.} vectors) between two sequences. It has been lately applied to the localization problem in \cite{schymura_exploiting_2021}, in the context of sound event detection and localization (SELD). \emph{Self-attention} is a particular variant of attention that analyses a single sequence, \emph{i.e.} it models the similarity of its elements. In \cite{phan_audio_2020}, the authors proposed to couple a CRNN with self-attention mechanism, also in order to improve the SELD baseline.

In this work, we are interested in adapting the CRNNs used for multiple speaker localization (thus, different from SELD). We propose to replace the recurrent layers by the encoder modules of the so-called \emph{Transformer} architecture \cite{vaswani_attention_2017}. Transformer uses standalone self-attention blocks, i.e.~it does not include recurrent layers at all. Originally applied to NLP tasks, it has since become very popular, as it can actually surpass the performance of classical RNNs. We thus propose to replace the bidirectional long short-term memory (BiLSTM) layers of a state-of-the-art CRNN \cite{grumiaux_improved_2021} by one or several self-attention encoders, depending on the model. As demonstrated on the simulated and real data, this modification actually improves the overall accuracy of the considered CRNN, while avoiding the recurrent layers and thus being better suited for parallel computing. We term these new architectures \emph{Self-Attentive multisource Localization in the Ambisonics Domain (SALAD) nets}.

\vspace{-0.3cm}
\section{Proposed Method}

In this section, we present the proposed SSL system, including input features, output format and network architecture.

\subsection{Input features}

We use the same input features as in \cite{perotin_crnn-based_2019,grumiaux_improved_2021}, namely the intensity vector from the first-order Ambisonics (FOA) representation of the audio signal. The Ambisonics format has proven to be well-suited for SSL \cite{perotin_crnn-based_2019, adavanne_direction_2018} as it provides a convenient way of representing the spatial properties of the soundfield. The FOA consists of the omnidirectional channel $W(t,f)$, and three figure-of-eight channels, aligned with Cartesian coordinate axes, denoted as $X(t,f)$, $Y(t,f)$ and $Z(t,f)$ ($t$ and $f$ denote the STFT time and frequency bins, respectively).\\footnote{We adopt the N3D Ambisonics normalization standard.}



We can then derive the expressions for the active and reactive intensity vectors in the FOA representation \cite{jarrett2017theory} (indexes $t$ and $f$ are omitted for concision, and $^*$ denotes the complex conjugate):
\begin{equation}
    \boldsymbol{I}_a = \begin{bmatrix} \operatorname{Re}\{WX^*\} \\ \operatorname{Re}\{WY^*\} \\ \operatorname{Re}\{WZ^*\} \end{bmatrix},
    \quad
    \boldsymbol{I}_r = \begin{bmatrix} \operatorname{Im}\{WX^*\} \\ \operatorname{Im}\{WY^*\} \\ \operatorname{Im}\{WZ^*\} \end{bmatrix}.
\end{equation}
For each time-frequency (TF) bin, the above active and reactive intensity vectors are concatenated to form a 6-channel vector
which is normalized by the sound power given by $|W(t,f)|^2 + \frac{1}{3}(|X(t,f)|^2+|Y(t,f)|^2+|Z(t,f)|^2)$. The resulting vector is reminiscent of the so-called \textit{Frequency Domain Velocity Vector} representation \cite{daniel_time_2020}. Finally, the input features are given as the collection of these vectors for all time-frequency bins, assembled into a tensor of size $N \times F \times 6$, with $N$ the number of frames and $F$ the number of frequency bins. In our experiments, the signals are sampled at $16$~kHz and we use a 1,024-point ($64$~ms) STFT (i.e. $F = 513$) with sinusoidal analysis window and $50$~\% overlap. The input sequences contains $25$ frames (representing $800$~ms of signal), hence the shape input feature is $25 \times 513 \times 6$.

\vspace{-0.2cm}
\subsection{Output}
\label{subsec:Output}

As in \cite{grumiaux_improved_2021}, we treat SSL as a classification problem. The 2D sphere is divided into a quasi-uniform grid of candidate DOA regions with elevation $\phi_i \in [-90,90]$ and azimuth $\theta^j_i \in [-180,180]$ given by:
\begin{equation}
    \begin{cases}
        \phi_i = -90 + \frac{i}{I} \times 180 & \text{with $i \in \{0, ..., I\}$} \\
        \theta_i^j = -180 + \frac{j}{J^i+1} \times 360 & \text{with $j \in \{0, ..., J^i\}$}, \\
    \end{cases}
\end{equation}
where $I = \lfloor \frac{180}{\alpha} \rfloor$ and $J^i = \lfloor \frac{360}{\alpha} cos \phi_i \rfloor$ with $\alpha$ the grid resolution in degrees.
Using classification-based localization, each region correspond to a class and the output vector $y$ is of size $C$ (which is the total number of classes). For example for the target, $y(c)=1$ if a source is present in a zone corresponding to class $c$, and $y(c)=0$ otherwise. We set $\alpha=10$\textdegree, resulting in a total of $C=429$ classes.

\vspace{-0.2cm}
\subsection{Network architecture}

\begin{figure}[t]
   \begin{subfigure}[c]{0.45\columnwidth}
        \includegraphics[width=\linewidth]{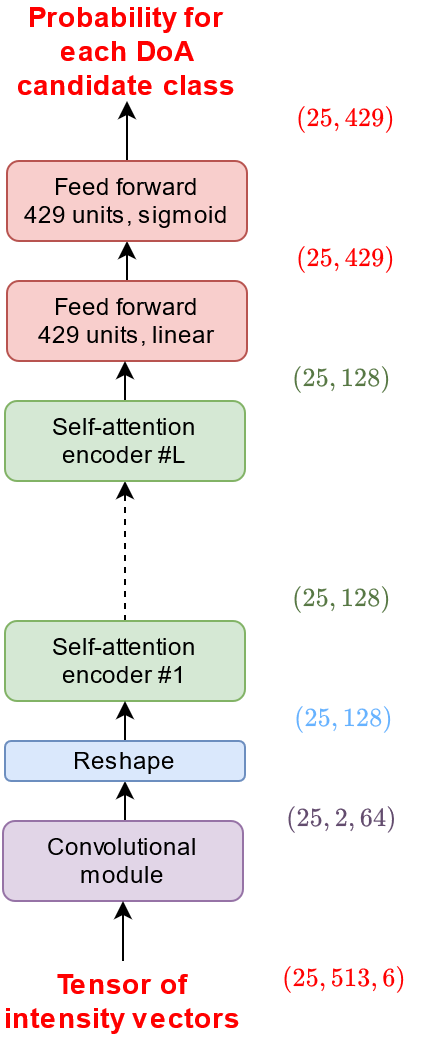}
    \end{subfigure}
    \begin{subfigure}[c]{0.45\columnwidth}
        \includegraphics[width=\linewidth]{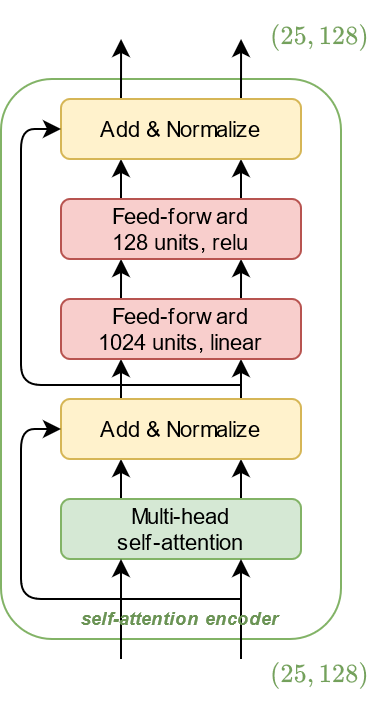}
    \end{subfigure}
    \caption{Proposed architecture based on convolutional layers and self-attention encoders. Left: overall architecture, with $L$ stacked self-attention encoders; Right: Detail of a self-attention encoder.}
    \vspace{-0.4cm}
    \label{fig:archi}
\end{figure}

As stated above, the proposed architecture is based on the architecture called Model~6-4 in \cite{grumiaux_improved_2021}, which is composed of a convolutional module followed by a recurrent module and ending with feed-forward layers for classification. 
Taking into account the limitations of the recurrent layers mentioned in the introduction, and the empirical observation that certain frames are more informative than others with regards to the SSL task \cite{perotin_crnn-based_2019}, we propose to replace the recurrent part of the model in \cite{grumiaux_improved_2021} with a self-attention-based encoder module. While recurrent layers only perform the analysis of temporal sequences in a sequential way, the use of self-attention allows for comparison of \textit{all}\footnote{In the present study, we do not use positional encoding \emph{c.f.} \cite{vaswani_attention_2017}.} frames with each other.
 Fig.~\ref{fig:archi} shows the diagram of the proposed neural network architecture, which is detailed in the remainder of this section.

\textbf{Convolutional module}: The $N \times F \times 6$ input features are first provided to a convolutional module identical to the one of Model 6-4 in \cite{grumiaux_improved_2021}, which was chosen due to its good trade-off between localization accuracy and model size. It is composed of 6 convolutional blocks, each block being made of two consecutive convolutional layers with 64 filters of size $3 \times 3$, followed by a max-pooling layer with various size on the second dimension. This module performs feature extraction by reducing the overall feature size  while keeping the temporal dimension ($N = 25$) of the input intact. After the convolutional module, the feature maps are concatenated together to form a 2D tensor of size $N \times G$ (reshaping module; here $G=128$).

\textbf{Self-attention module}: The sequence of reshaped features goes into a stack of $L$ self-attention encoders. Each self-attention encoder follows the structure shown in the right side of Fig.~\ref{fig:archi}, which is the encoder of the Transformer architecture \cite{vaswani_attention_2017}. It is composed first of a multi-head self-attention layer (classical multi-head or cross multi-head according to the experiment, detailed below) which produces an encoded feature for each feature of the input $N$-sequence. Each encoded feature is then added to the corresponding input feature with a residual connection, and the sum is normalized. After that, each sequence item goes through the same two-layer feed-forward module, the first layer being linear and the second layer having a ReLU activation. Then again, a residual connection and normalization are applied. The output of such a self-attention encoder has the same dimension as its input (here $N \times G$).

Self-attention has been originally proposed for sequential data, and works as following (see \cite{vaswani_attention_2017} for more details). For each vector $\mathbf{x}_i$ of the input sequence of length $N$, three vectors $\mathbf{q}_i$, $\mathbf{k}_i$ and $\mathbf{v}_i$ (usually labelled as \textit{query}, \textit{key} and \textit{value}, respectively) are computed by multiplying $\mathbf{x}_i$ with learnt matrices $\mathbf{W}^Q$, $\mathbf{W}^K$ and $\mathbf{W}^V$. Then, for each vector $\mathbf{x}_i$, an encoded vector $\mathbf{z}_i$ is calculated via weighted sum of all $\{\mathbf{v}_j\}$: $\mathbf{z}_i = \sum_{j=1}^{N} s_{ij} \mathbf{v}_j$, where $s_{ij}$ is the score of $\mathbf{x}_i$ against $\mathbf{x}_j$. The score $s_{ij}$ is calculated as the softmax of a scaled dot product of $\mathbf{q}_i$ and $\mathbf{k}_j$, so that the sum of $s_{ij}$ over $j$ is equal to $1$:
\begin{equation}
    s_{ij} = \text{softmax}\Big(\frac{\mathbf{q}_i \cdot \mathbf{k}_j}{\sqrt{G}}\Big)
\end{equation}
In summary, self-attention applied on the temporal axis computes at each time index $i$ an output vector $\mathbf{z}_i$ that takes into account the input vector $\mathbf{x}_i$ at time $i$ and its dependencies (the scores) with the other (past and future) vectors $\mathbf{x}_j$ in the sequence. Note that $\mathbf{x}_i$ and $\mathbf{z}_i$ are vectors of the same size. 


In the Transformer paper \cite{vaswani_attention_2017}, the authors also introduce the concept of multi-head self-attention, which consists of learning $H$ independent sets of matrices $\mathbf{W}^Q_h$, $\mathbf{W}^K_h$ and $\mathbf{W}^V_h$, with $h \in [1,H]$ and $H$ the number of heads, thus leading to $H$ vectors $\mathbf{q}_{ih}$, $\mathbf{k}_{ih}$ and $\mathbf{v}_{ih}$ for each input $\mathbf{x}_i$. The score are computed independently for each head $h$ :
\begin{equation}
    s_{ijh} = \text{softmax}\Big(\frac{\mathbf{q}_{ih} \cdot \mathbf{k}_{jh}}{\sqrt{G}}\Big)
\end{equation}
In this case, a multi-head version of $\mathbf{z}_{i}$ is calculated as a weighted sum of all $\mathbf{v}_{jh}$ : $\mathbf{z}_{ih} = \sum_{j=1}^{N} s_{ijh} \mathbf{v}_{jh}$. The $\mathbf{z}_{ih}$ are finally concatenated in the head dimension and another learnt matrix $\mathbf{W}^O$ is used to output one vector $\mathbf{z}_i$ for each input vector $\mathbf{x}_i$ as in single-head self-attention. This multi-head aspect provides more flexibility in the self-attention mechanism.

We also evaluate the use of a more general way of computing the multi-head attention scores, where a score is calculated independently for each head pair $(h,h')$ (while in \cite{vaswani_attention_2017} the scores are calculated head-wise) :
\begin{equation}
    s_{ijhh'} = \text{softmax}\Big(\frac{\mathbf{q}_{ih} \cdot \mathbf{k}_{jh'}}{\sqrt{G}}\Big)
\end{equation}
Then $\mathbf{z}_{ih}$ is obtained as a weighted sum on $j$ and $h'$ : $\mathbf{z}_{ih} = \sum_{j=1}^{N} \sum_{h'=1}^{H} s_{ijhh'} \mathbf{v}_{jh'}$. We call this method cross multi-head (CMH) self-attention to distinguish it from original multi-head self-attention \cite{vaswani_attention_2017}. In our experiments , $H$ is a hyperparameter.


\textbf{Feed-forward fully-connected module}: At the end, each vector of the output sequence of the self-attention module is sent to the same two-layer feed-forward network, with $429$ units in each layer. The first layer is linear, and the second one has the sigmoid activation, to provide output values homogeneous to probability of having a source present in the corresponding DOA region (see Section~\ref{subsec:Output}).

\vspace{-0.2cm}
\section{Experiments}

\vspace{-0.1cm}
\subsection{Data}

To generate data for training and testing, we used the same methodology as in \cite{grumiaux_improved_2021}. Beforehand, we adapted the spatial room impulse response (SRIR) generator of \cite{habets_room_2006} (based on the image-source method \cite{allen_image_1979}) to synthesize FOA impulse responses for many different room configurations: room lengths, widths and heights randomly drawn in $[2, 10]$~m, $[2, 10]$~m and $[2, 3]$~m, respectively; RT60s lie in $[200,800]$~ms; microphone and source positions are randomly picked in the room with the microphone at least $0.5$~m from the walls. These synthetic SRIRs are used for the training dataset as well as for the first test dataset. For a second more realistic test dataset, we also recorded SRIRs in our acoustic lab (RT60 $\approx$ 500~ms), using all combinations from 36 microphone positions and 16 static loudspeaker positions. We generated microphone signals by convolving the SRIRs (either real or synthetic) with TIMIT speech signals \cite{garofolo_timit_1993}. For each room configuration, we created signals corresponding to a single source, 2- and 3- source mixture. The signal-to-interference ratio between speakers is randomly drawn between 0 and 10~dB. Some diffuse babble noise, obtained by averaging the late reverberation part of two random synthetic RIRs, was added to those mixtures with a random SNR between 0 and 20~dB. The resulting training dataset was composed of 247,400 sequences for each number of speakers (1 to 3), i.e. a total of 772,200 training sequences (about 172 hours of signals). Each test set (synthetic and real data) contains 1152 sequences for each number of speakers, corresponding to about 45 minutes of signals.

\vspace{-0.1cm}
\subsection{Training procedure \& parameters}

We designed and trained the proposed model using Keras framework on Nvidia GTX 1080 GPUs. We use the Nadam optimizer \cite{dozat_incorporating_2016} with default parameters. Early stopping was applied with a patience of 20 epochs by monitoring the accuracy on a validation set, and the learning rate was divided by two when the validation accuracy was not improving after 10 epochs.

\vspace{-0.1cm}
\subsection{Configurations}

We trained and tested several variants of the proposed model to evaluate the capacity of the self-attention module to replace the recurrent layers for SSL. In the presented results, we use the naming convention ``\textbf{$X$-$N_{\mathrm{enc}}$enc-$H$H}'', where $X$ is \textit{MH} or \textit{CMH} whether multi-head or cross multi-head is used, respectively, $N_{\text{enc}}$ denotes the number of encoders, and $H$ is the number of attention heads. First, we tested $H \in \{1, 2, 3, 10\}$ using classical multi-head attention. Then, we evaluated the benefit of using cross multi-head self-attention for $H \in \{3, 10\}$. Finally, we investigated the use of two cross multi-head self-attention encoders instead of 1, with $H \in \{5, 10\}$. 

\vspace{-0.1cm}
\subsection{Metrics, evaluation and baseline}

During the inference, the output of the neural network is averaged over the frames, meaning that we end up with one probability distribution for the whole input sequence. Knowing the number of sources $S$ in the test signal, we extract the $S$ highest peaks in the probability distribution, in which a peak represents a local maximum within the spherical geometry.
The DOA estimation task is evaluated by computing the sequence-wise accuracy, i.e.~the percentage of sequences with a predicted DOA angular error lower than a given tolerance (the higher the better). Since the minimal angular separation between two points in our grid is 7\textdegree, we set tolerance threshold to 10\textdegree~or 15\textdegree. We also evaluated the performance with the mean and median angular error, averaged on all test sequences (the lower the better), thus we indicate the standard deviation as well. As the proposed architectures are based on the CRNN model called Model~6-4 in \cite{grumiaux_improved_2021}, we use this model as a common baseline, referred as \textit{Baseline (CRNN)}. A DNN-free baseline is also referred as \textit{Baseline (TRAMP)}: this method is based on the histogram - over all frames and frequencies of the sequence - of the DOAs derived from the pseudointensity vector; each DOA is weighted by an ad-hoc plane-wave indicator (see \cite{kitic_tramp:_2018} for more details).

\vspace{-0.1cm}
\subsection{Results}
\label{sec:results}

\begin{table*}[h!]
\resizebox{2.\columnwidth}{!}{%
\begin{tabular}{|c|ccccc|ccccc|ccccc|}
\hline
\multirow{2}{*}{\textbf{Model}} & \multicolumn{5}{c|}{\textbf{1 source}}                                                      & \multicolumn{5}{c|}{\textbf{2 sources}}                                                     & \multicolumn{5}{c|}{\textbf{3 sources}}                                                     \\
                                & \textbf{Acc. \textless{}10°} & \textbf{Acc. \textless{}15°} & \textbf{Mean} & \textbf{Med.} & \textbf{Std.} & \textbf{Acc. \textless{}10°} & \textbf{Acc. \textless{}15°} & \textbf{Mean} & \textbf{Med.} & \textbf{Std.} &\textbf{Acc. \textless{}10°} & \textbf{Acc. \textless{}15°} & \textbf{Mean} & \textbf{Med.} &\textbf{Std.} \\ \hline
Baseline (TRAMP)                 & 43.7                         & 64.2                         & 16.7           &  11.5  & 20.3        & 27.9                         & 39.9                         & 44.0           & 21.9   & 44.6       &  22.4                      &  33.6                        &  48.4        & 30.0  & 44.9      \\
Baseline (CRNN)                 & \textbf{98.6}                         & 99.7                         & \textbf{4.4}           & \textbf{4.1} & 4.6          & 88.3                         & 93.3                         & 7.7           & \textbf{4.7} & 19.4           & 74.7                         & 83.4                         & 12.8          & 5.9  & 29.7         \\
MH-1enc-1H                      & 98.3                         & 99.7                         & 4.5           & 4.2           & 4.5 & 87.9                         & 93.6                         & 7.9           & 4.9 & 13.8           & 71.6                         & 81.5                         & 13.5          & 6.5 & 21.1          \\
MH-1enc-2H                      & 98.1                         & 99.7                         & 4.7           & 4.2    & 5.2       & 87.1                         & 93.0                         & 8.7           & 4.9 & 15.3          & 72.2                         & 80.6                         & 15.8          & 6.2  & 23.9         \\
MH-1enc-3H                      & 98.1                         & 99.6                         & 4.7           & 4.2   & 5.5        & 87.7                         & 92.7                         & 8.8           & 4.8 & 16.0          & 72.9                         & 80.7                         & 16.2          & 6.0 & 25.1          \\
MH-1enc-10H                     & 98.5                         & 99.6                         & 4.5           & \textbf{4.1}  & 5.5         & \textbf{90.4}                         & 94.5                         & 7.4           & \textbf{4.7} & 15.4           & 77.3                         & 84.7                         & 12.3          & \textbf{5.6} & 24.2          \\
CMH-1enc-3H                      & 98.5                         & \textbf{99.8}                         & \textbf{4.4}           & \textbf{4.1} & 5.1          & 89.3                         & 94.1                         & 7.6           & 4.8 & 15.1          & 75.9                         & 84.1                         & 12.7          & 5.9 & 23.6          \\
CMH-1enc-10H                     & 98.4                         & 99.5                         & 4.5           & \textbf{4.1}  & 5.1         & 89.9                         & 94.5                         & \textbf{6.8}           & \textbf{4.7}  & 14.6         & \textbf{78.2}                         & \textbf{85.7}                         & \textbf{11.3}          & \textbf{5.6}  & 23.0         \\
CMH-2enc-5H                      & 98.5                         & 99.6                         & 4.6           & 4.2  & 4.9         & \textbf{90.4}                         & 94.7                         & 7.0           & \textbf{4.7} & 14.3           & 75.6                         & 84.2                         & 12.4          & 6.0 & 22.8          \\
CMH-2enc-10H                     & 97.9                         & 99.3                         & 4.8           & 4.2  & 5.0         & 88.7                         & \textbf{94.9}                         & 7.3           & 5.0 & 14.1          & 72.8                         & 83.7                         & 13.0          & 6.5 & 22.5          \\ \hline
\end{tabular}}
\vspace{-0.1cm}
\caption{SSL results on the test dataset generated with synthetic SRIRs (best results are in bold).}
\label{tab:resultsSynthetic}
\end{table*}

\begin{table*}[h!]
\resizebox{2.\columnwidth}{!}{%
\begin{tabular}{|c|ccccc|ccccc|ccccc|}
\hline
\multirow{2}{*}{\textbf{Model}} & \multicolumn{5}{c|}{\textbf{1 source}}                                                      & \multicolumn{5}{c|}{\textbf{2 sources}}                                                     & \multicolumn{5}{c|}{\textbf{3 sources}}                                                     \\
                                & \textbf{Acc. \textless{}10°} & \textbf{Acc. \textless{}15°} & \textbf{Mean} & \textbf{Med.}&\textbf{Std.} & \textbf{Acc. \textless{}10°} & \textbf{Acc. \textless{}15°} & \textbf{Mean} & \textbf{Med.}&\textbf{Std.} & \textbf{Acc. \textless{}10°} & \textbf{Acc. \textless{}15°} & \textbf{Mean} & \textbf{Med.}&\textbf{Std.} \\ \hline
Baseline (TRAMP)                 &  52.9                        & 72.8                         & 14.5           &  9.4  & 19.2        &  30.5                        & 42.9                         &  37.0          &  19.1  &  38.6      &  25.1                      &  35.6                        &  43.0        & 26.0  & 41.1      \\
Baseline (CRNN)                 & \textbf{79.0}                         & \textbf{93.7}                         & 7.6           & \textbf{6.1} & 10.3          & 68.2                         & 84.7                         & 11.9          & \textbf{7.2} & 24.8           & 56.8                         & 73.3                         & 17.3          & 8.7 & 32.9          \\
MH-1enc-1H                      & 77.0                         & 93.5                         & \textbf{7.5}           & 6.2 & 6.5          & 67.4                         & 83.5                         & 11.5          & \textbf{7.2} &15.8           & 53.8                         & 69.9                         & 18.9          & 9.1 & 26.1           \\
MH-1enc-2H                      & 76.8                         & 93.4                         & 7.6           & 6.3  & 6.8         & 67.9                         & 83.8                         & 12.7          & 7.5 & 18.3          & 53.6                         & 68.5                         & 22.5          & 9.1  & 29.2         \\
MH-1enc-3H                      & 76.2                         & 92.7                         & 8.1           & 6.3 & 8.1          & 67.4                         & 84.9                         & 12.3          & 7.3  & 18.4         & 54.6                         & 68.3                         & 21.8          & 9.1 & 29.4          \\
MH-1enc-10H                     & 77.3                         & 93.0                         & 8.3           & 6.2 & 9.0          & 68.2                         & 86.3                         & 11.2          & 7.3 & 18.0          & 57.8                         & 74.0                         & 16.9          & \textbf{8.5} & 28.2          \\
CMH-1enc-3H                      & 77.5                         & 92.6                         & 8.0           & 6.3 & 9.1          & 68.6                         & 85.6                         & 10.7          & 7.3 & 17.3          & 56.4                         & 72.3                         & 18.2          & 8.9 & 27.8          \\
CMH-1enc-10H                     & 77.0                         & 92.6                         & 8.0           & 6.2 & 9.1          & 68.6                         & 85.8                         & 10.5          & 7.3  & 16.7         & \textbf{58.2}                         & \textbf{74.8}                         & \textbf{15.2}          & \textbf{8.5}  & 26.8         \\
CMH-2enc-5H                      & 75.7                         & 92.6                         & 8.4           & 6.3 & 9.4          & \textbf{70.0}                         & \textbf{87.1}                         & \textbf{10.4}          & \textbf{7.2} & 16.4          & 57.7                         & 74.2                         & 16.3          & 8.6  & 26.3         \\
CMH-2enc-10H                     & 75.7                         & 91.1                         & 8.8           & 6.2 & 9.9          & 69.0                         & 86.9                         & 10.6          & \textbf{7.2} & 16.2          & 56.3                         & 73.3                         & 17.1          & 8.9 & 26.0          \\ \hline
\end{tabular}}
\vspace{-0.1cm}
\caption{SSL results on the test dataset generated with real SRIRs (best results are in bold).}
\vspace{-0.2cm}
\label{tab:resultsReal}
\end{table*}

\begin{table}[h]
\centering
\begin{tabular}{|c|cc|}
\hline
\textbf{Model}  & \textbf{real-time \%} & \textbf{\# parameters} \\ \hline
Baseline (CRNN) & 437  & 913,907    \\
MH-1enc-1H      & 244  & 796,125    \\
MH-1enc-2H      & 244  & 862,045    \\
MH-1enc-3H      & 244  & 927,965    \\
MH-1enc-10H     & 244  & 1,389,405  \\
CMH-1enc-3H      & 244  & 927,965    \\
CMH-1enc-10H     & 244  & 1,389,405  \\
CMH-2enc-5H      & 281  & 1,653,341  \\
CMH-2enc-10H     & 281  & 2,312,541  \\ \hline
\end{tabular}
\caption{Real-time percentage for inference and number of parameters for the different models (in our experiments, frame length = 0.032~s).}
\vspace{-0.5cm}
\label{tab:times}
\end{table}

The results obtained on the synthetic and real SRIRs test datasets are shown in Table~\ref{tab:resultsSynthetic} and Table~\ref{tab:resultsReal}, respectively. First, we can see that all neural-based systems largely surpass the DNN-free baseline from \cite{kitic_tramp:_2018}. Second, we can see that the performance of all the tested self-attention-based neural networks are roughly similar to the performance of the baseline CRNN, sometimes a bit below and sometimes better, depending on the configuration (we detail below). This shows that it is possible to replace the recurrent layers with self-attention encoders without losing in performance. Importantly, we can see in Table~\ref{tab:times} that the gain in inference time over the baseline is quite significant. The median inference time using self-attention-based networks is 44\% lower than for the baseline (when using one self-attention encoder; it is  33\% lower when using 2 encoders). We also see that this inference time is not correlated with the number of parameters, but rather with the number of self-attention encoders, which is to be expected.

Generally, the MH attention-based networks with 1, 2 or 3 heads perform slightly below the baseline. However, the MH self-attention network with 10 attention heads outperforms the baseline in the multi-source configurations: for example, on 3-source mixtures with synthetic SRIRs, the accuracy (< 10\textdegree) is 78.2~\% vs 74.7~\% for the baseline, and the mean angular error is lower by 1.5\textdegree. The same trend can be observed for the real SRIRs configuration, though with a more moderate performance gain. This suggests that adding more heads to the self-attention mechanism may be particularly beneficial for multi-source localization. 

The results for the models with CMH self-attention are also globally better than the baseline (in the multi-source configurations), and also better than MH self-attention when comparison is made with the same number of heads. CMH self-attention leads to an improvement for both 3 and 10 heads over MH self-attention, especially in terms of mean and median angular error. For example, going from MH to CMH self-attention with 10 heads reduce the mean error by 0.6\textdegree~and 1\textdegree~for 2-source and 3-source mixtures (synthetic SRIRs), respectively, and by 0.7\textdegree~and 1.7\textdegree~for 2-source and 3-source mixtures (real SRIRs). The accuracy is also improved, up to 78.2\% (< 10\textdegree) for synthetic SRIRs and 58.2\% for real SRIRs, representing a performance gain of 3.5\% and 1.4\% over the baseline, respectively. Allowing the score calculation across all heads seems to provide more flexibility to the self-attention mechanism thus leading to more robust multi-source localization.

Finally, the results of Models CMH-2enc-5H and CMH-2enc-10H shows what happens when we stack two CMH self-attention encoders (for the same number of heads). We can see that the performance of model CMH-2enc-5H model is better than the one of model CMH-1enc-10H for 2-source mixtures (for both synthetic or real SRIRs), but lose performance on 3-source signals. The mean error increases by 1.1\textdegree. When dealing with 2 encoders with 10 attention heads each, the performance gets even worse, somehow going back near to the baseline performance. The shows that adding more encoders does not necessarily improve the performance, possibly because the complexity of the network becomes too high for this task or for the amount of training data available.

\vspace{-0.3cm}
\section{Conclusion}
\vspace{-0.1cm}
We have presented a novel multi-head self-attentive neural network for SSL of up to 3 speakers. We first showed that self-attention encoders are suitable for replacing recurrent layers of a state-of-the-art CRNN without losing in localization performance, while saving computations and inference time. We also compared the performance for different configurations of the self-attention module. Overall, the best performance was obtained with a cross multi-head self-attention model with one 10-head encoder, which clearly outperformes the baseline state-of-the-art CRNN in the multi-source configuration (again for a lower computation cost). Future work will further investigate the performance of these models, e.g.~why the increase of the number of stacked encoders does not improve the performance in our experiments. 

\vspace{-0.2cm}
\bibliographystyle{IEEEtran}
\bibliography{ref}
%
%
%
%
%
%
%
%
%

\end{sloppy}
\end{document}